\documentclass[12pt,preprint]{aastex} 

\IfFileExists{srcltx.sty}{\usepackage[active]{srcltx}} 

\slugcomment{Submitted to the {\it Astrophysical Journal}}
\slugcomment{UCLA/12/TEP/00}
\shorttitle{Sterile Neutrinos in Willman 1 from XMM}
\shortauthors{Loewenstein and Kusenko}

\begin{document}
\title{Dark Matter Search Using {\em XMM-Newton} Observations of
 Willman 1}

\author{Michael Loewenstein\altaffilmark{1,2,3}, and Alexander
Kusenko\altaffilmark{4,5}}

\altaffiltext{1}{Department of Astronomy, University of Maryland,
 College Park, MD.} 
\altaffiltext{2}{Center for Research and
 Exploration in Space Science and Technology} 
\altaffiltext{3}{X-ray Astrophysics Laboratory, Mail Code 662, NASA
 Goddard Space Flight Center, Greenbelt, MD 20771, USA}
\altaffiltext{4}{Department of Physics and Astronomy, University of
 California, Los Angeles, CA 90095-1547, USA}
\altaffiltext{5}{Institute for the Physics and Mathematics of the
 Universe, University of Tokyo, Kashiwa, Chiba 277-8568, Japan}


\begin{abstract}
We report the results of a search for an emission line from
radiatively decaying dark matter in the ultra-faint dwarf spheroidal
galaxy Willman 1 based on analysis of spectra extracted from {\it
 XMM-Newton} X-ray Observatory data. The observation follows up our
analysis of {\it Chandra} data of Willman 1 \citep{lk10} that resulted
in line flux upper limits over the {\it Chandra} bandpass and evidence
of a 2.5 keV feature at a significance below the 99\% confidence
threshold used to define the limits. The higher effective area of the
{\it XMM-Newton} detectors, combined with application of recently
developing methods for extended-source analysis, allow us to derive
improved constraints on the combination of mass and mixing angle of
the sterile neutrino dark matter candidate. We do not confirm the
{\it Chandra} evidence for a 2.5 keV emission line.
\end{abstract}



\section{Introduction}

Sterile neutrinos represent a plausible dark matter candidate amenable
to observation via X-ray spectroscopy of locations in the universe
where the dark matter surface density is high. Sterile neutrinos,
gauge-singlet fermions that emerge in extensions of the Standard Model
of particle physics that explain ordinary neutrino masses, may
comprise some or all of the dark matter if the Majorana masses are
below the electroweak scale \citep{kus09}. If their mass lies in the
1--30 keV range, non-resonant oscillations produce sterile neutrinos
at an abundance comparable to that inferred for dark matter
\citep{dw94} -- although several other production channels have been
suggested \citep{kus09}. Sterile neutrinos in this mass range may
explain the observed velocities of pulsars by anisotropic emission of
sterile neutrinos from a cooling neutron star born in a supernova
explosion \citep{ks97,fkmp03} and can facilitate early star formation
\citep{kus09}. As a form of warm dark matter (WDM), sterile neutrinos
\citep{afp01,pet08,boy08,bw11,dun11} may resolve some of the
discrepancies (Ferrero et al. 2011, and references therein) between
Cold Dark Matter (CDM) models \citep{lov11,mfl12} and the observed
structure in the universe.  A number of particle physics models have
been proposed to accommodate a sterile neutrino with the requisite
mass and mixing, including $\nu$MSM~\citep{as05,brs09}, split seesaw
\citep{kty10}, Higgs singlet ~\citep{kus06,pk08}, and other models.
For example, the split seesaw mechanism \citep{kty10} with one extra
dimension causes the mass and mixing of a sterile neutrino originating
at a high scale to be exponentially suppressed in the low-energy
effective theory, which makes it a good dark matter candidate, while
preserving both the standard seesaw explanation for the neutrino
masses and the explanation of the matter-antimatter asymmetry of the
universe by leptogenesis~\citep{fy86}.

The one-loop decay of relic keV sterile neutrinos into an active
neutrino and photon with energy $E_\gamma = m_{\rm st}c^2/2$ produces
an X-ray emission line with width corresponding to the velocity
dispersion of the dark matter distribution within the observed solid
angle.  While the prospects of discovering a dark-matter sterile
neutrino in a laboratory present a daunting challenge
\citep{ak10,bs07}, the narrow decay line allows one to search for
relic sterile neutrinos using X-ray telescopes~\citep{aft01}.  The
high dark matter concentrations of dwarf spheroidal galaxies make them
prime targets for dark matter searches, with the additional advantage
of the total absence of competing intrinsic X-ray sources. Moreover,
as the most dark matter dominated objects known they are not subject
to systematic uncertainties that derive from decomposing the mass into
baryonic and non-baryonic components -- although assumptions about
membership and dynamical equilibrium come into play in converting
measured velocity dispersion profiles into dark matter mass
estimates. We initiated a dedicated search for sterile neutrinos with
the {\em Suzaku} X-ray telescope, placing new limits on sterile
neutrinos over the 1--20 keV mass range from XIS spectra of the Ursa
Minor dwarf spheroidal galaxy \citep{lkb09}.

We continued the search with {\it Chandra} observation of the
ultra-faint dwarf spheroidal Willman 1 (Loewenstein \& Kusenko 2010,
\apj, 700, 426; hereafter LK10), considered at the time to be a
particularly compelling target for dark matter searches (though its
status subsequently changed somewhat; see below). 99\% confidence line
flux upper limits over the 0.4-7 keV {\it Chandra} bandpass were
derived and mapped to an allowed region in the sterile neutrino
mass-mixing angle plane consistent with the {\it Suzaku}
constraints. In addition we reported evidence for an emission line
with flux below this threshold from radiative decay of a 5 keV sterile
neutrino with a mixing angle in the narrow range where oscillations
produce all of the dark matter and for which sterile neutrino emission
from the cooling neutron stars can explain pulsar kicks (LK10). This
tentative result is best confirmed (or refuted) by utilizing the large
effective area of the {\it XMM-Newton} EPIC detectors; and, we were
awarded observing time on Willman 1 for this purpose. Here we present
results of our analysis of these data, deriving more stringent
constraints on sterile neutrino parameters and failing to confirm the
best-fit estimate of the {\it Chandra} 2.5 keV line flux.

\section{Data Analysis}

\subsection{Observation and Data Processing}

Willman 1 was observed with {\it XMM-Newton} in three segments in late
October 2010: ObsID 0652810101 (for 29356 s on 10/22), ObsID
0652810301 (36054 on 10/25), and ObsID 0652810401 (36218 s on
10/31). In addition, a region offset by 1 degree was observed (ObsID
0652810201; 33856 s on 10/30). The last was primarily meant as a
control in the event of a positive emission line detection; and, as
none was found we will not consider these data further (although they
were reduced and analyzed in parallel to the on-source data). We
utilize data from the two EPIC-MOS (hereafter, EMOS = EMOS1 + EMOS2),
as well as the EPIC-PN (hereafter, EPN), CCD detectors. Data
reduction, background modeling, and spectral extraction are conducted
with the {\it XMM-Newton} Extended Source Analysis Software (XMM-ESAS)
-- part of the XMM-Newton Science Analysis System (SAS v11.1 is used
throughout) -- and methods as detailed in \cite{ks08} and
\cite{sno08}. Although various extended source analysis techniques are
adopted in X-ray dark matter searches, we are the first to apply this
specfic methodology optimized for extended low surface brightness
X-ray emission. A summary of our particular application of these
techniques follows; further details on the approach and individual
procedures and may be found in the XMM-ESAS
``cookbook''\footnote{http://heasarc.gsfc.nasa.gov/docs/xmm/xmmhp\_xmmesas.html}.

In order to apply the latest calibration files and products, rather
than use pipeline-processed data, the unprocessed event files are
reduced to create filtered event files using the composite tools {\it
 emchain} and {\it mos-filter} ({\it epchain} and {\it pn-filter})
for EMOS (EPN) datasets. This results in the removal of time intervals
of prominent soft proton (SP) flaring -- identified by hard (2.5-12
keV) energy band background count rates deviating by more than
$1.5\sigma$ from the mean of a Gaussian distribution that
characterizes the count rate histogram. Out-of-time (OOT) EPN events
that correspond to photons registered during CCD readout are processed
in parallel to ordinary EPN events for subsequent correction (see
below). Exposure times for the final cleaned event files are shown in
Table 1. Primarily due to flaring, data from $\sim$one-third of the
EMOS, and $\sim 60$\% of the EPN, exposures are excluded in the end.

\begin{deluxetable}{rlll}
\tabletypesize{\scriptsize}
\tablewidth{0pt}
\tablecaption{Final Good Time Intervals in ks}
\tablehead{\colhead{ObsId} & \colhead{EMOS1} & \colhead{EMOS2} & \colhead{EPN}}
\startdata 
0652810101 & 15.37 & 19.21 & 9.53\\
301 & 21.86 & 23.13 & 15.47\\
401 & 27.30 & 28.43 & 16.16\\ \hline
total & 64.53 & 70.70 & 41.16\\ \hline
\enddata 
\end{deluxetable}

\subsection{The Extraction of Source and Background Spectra}

We exclude background point source as follows. Source detection is
conducted with the SAS v11 composite {\it edetect\_chain} tool as a
standalone procedure and as incorporated in the XMM-ESAS {\it cheese}
procedure. These are used in the construction of region masks in
detector coordinates for use in spectral extraction. We considered
masks that (a) minimize the excluded regions by restricting only the
cores (the inner $8''$ in radius) of the brightest sources (those with
likelihood$>40$), (b) that minimize contamination from point sources
($32''$ radii for sources detected with likelihood$>10$), as well as
(c) an intermediate choice ($16''$ radii for sources detected with
likelihood$>10$). We also considered three spectral extraction
regions. The first corresponds to $1.7\times$ the half-light radius
($1.7\times 25$ pc $\equiv 232''$ at the 38 kpc Willman 1 distance)
within which the mass is particularly well-determined \citep{ae11} and
for which we apply mask ``a'' above, the second to 150 pc ($818''$;
mask ``b''), and the third to 100~pc ($545''$; mask ``c'') -- the mass
distribution is estimated within 100~pc in \cite{str08}. The latter
two are of the same order as the optical extent of Willman 1, and may
be compared to the tidal radius of 930 pc (Section 3) and a virial
radius on the $10^9~\rm M_{\odot}$ mass-scale of $\sim 25$ kpc.

Files were processed with $FLAG == 0$ and screened to retain only
those events with $PATTERN<=12$ (four-or-fewer pixel events) for EMOS,
and $PATTERN<=4$ (single and double pixel events) for EPN, spectra as
a means to exclude non-X-ray events. The XMM-ESAS {\it mos-spectra}
and {\it mos-back} ({\it pn-spectra} and {\it pn-back}) procedures are
used to extract EMOS (EPN) source and quiescent particle background
(QPB) spectra, and to compute spectral response files. This method
utilizes filterwheel-closed data, data from the unexposed corners of
archived {\it XMM-Newton} observations, and ROSAT All-Sky Survey
(RASS) data to produce a model QPB spectrum. Chips found to be in
anomalous states with elevated low-energy background according to the
XMM-ESAS criteria (EMOS1-ccd4 and -ccd5 for ObsID 0652810301 and ObsID
0652810401, EMOS2-ccd5 for ObsID 0652810101) are excluded, and
EMOS-1ccd6 was no longer operating at the time of these
observations. The detector extraction areas are calculated using the
{\it proton\_scale} task. The spectral extraction procedure generates
instrument response matrices in the form of separate redistribution
matrix ({\bf rmf}) and effective area function ({\bf arf}) files,
taking into account the extended nature of the emission.

\subsection{Spectral Analysis}

\subsubsection{The Baseline Spectral Model}

The XMM-ESAS procedures create model QPB background spectra
appropriate to the conditions during the observation under
consideration; however, a number of internal and astrophysical
background components remain that must be included in any spectral
model. This approach is both more accurate and more conservative than
one where blank-sky background is subtracted, given the directional
dependence of the intensity and spectral shape of the cosmic
contribution, and the time dependence of the particle and instrumental
backgrounds \citep{ks08,sno08,lk10}. Moreover, in the event that there
is dark matter line emission the ``blank-sky'' includes a contribution
to the signal that ought not be subtracted.

The instrumental background includes a number of fluorescent
lines. The strongest of these -- the 1.49 keV Al $K\alpha$ (EMOS and
EPN) and 1.75 keV Si $K\alpha$ (EMOS only) features are sufficiently
variable that the XMM-ESAS model QPB spectrum includes a smooth bridge
over the relevant spectral region, and the lines must be included in
spectral modeling in the form of narrow Gaussians. Other problematic
EPN lines are avoided by restricting the bandpass to $<7$ keV.

Even after filtering, contamination from residual SP flaring and solar
wind charge exchange (SWCX) are commonly present. These are modeled
using a broken power-law and a pair of unresolved emission lines at
0.56 keV (OVII) and 0.65 keV (OVIII) that dominate SWCX spectra,
respectively. Since energy from the SP component is directly deposited
in the detectors, diagonal response files are used for this component
in place of the telescope responses described above.

The astrophysical X-ray background includes multiple components that
we model with thermal plasma ({\bf apec}) models to account for the
emission from the Local Hot Bubble (LHB) and the Milky Way Halo (MWH),
and a power-law for the unresolved point sources mostly originating
from background AGN (CXB). The elemental composition in the {\bf apec}
components are set at their solar abundances as defined in
\cite{agss09}. We use the {\tt HEASARC} background
tool\footnote{http://heasarc.gsfc.nasa.gov/cgi-bin/Tools/xraybg/xraybg.pl}
to extract the ROSAT All-Sky Survey (RASS) spectrum from a 1$\arcdeg$
radius aperture centered on the position of Willman 1 for use in
constraining the parameters of these components.

\subsubsection{Fitting Techniques and Best-Fit Baseline Models}

Spectra are fitted using {\sc Xspec} version
12.7\footnote{http://heasarc.gsfc.nasa.gov/docs/xanadu/xspec/}. The
final total spectral model we adopt may be expressed, in {\sc Xspec}
notation, as {\bf bknpower + gaussian + gaussian + constant $\times$
 constant $\times$ (gaussian + gaussian + apec + (gaussian + apec +
 apec + constant $\times$ powerlaw) $\times$ TBabs)}, where {\bf
 bknpower} represents the residual SP component (with its distinct
diagonal response matrix), the first pair of {\bf gaussians} the Al
$K\alpha$ and Si $K\alpha$ instrumental lines, and the second pair of
{\bf gaussians} the SWCX emission. The remaining astrophysical
background consists of the unabsorbed {\bf apec} LHB and absorbed
(two-temperature) {\bf apec + apec} MWH, as well as the absorbed {\bf
 powerlaw} CXB emission. The {\bf constant} factor multiplying the
CXB represents the departure of the CXB intensity from its all-sky
average \citep{kush02} due to cosmic variance and the fact that the
resolved portion of the CXB is removed. The Tuebingen-Boulder ISM
absorption model \citep{wam00}, {\bf tbabs}, is applied with the
column density fixed at the Galactic value of $1.13\times 10^{20}$
cm$^{-2}$ \citep{dl90}. The two multipliers that act on all of the sky
components, {\bf constant} $\times$ {\bf constant}, represent the size
of the spectral extraction region (see above) so that all
normalizations correspond to fluxes per solid angle, and a factor to
account for any detector calibration offset.

There are datasets from three detectors for each of the three
on-source ObsIDs. We initially analyze each dataset separately. As is
usually the case, we find that the EMOS1 and EMOS2 spectra are very
similar for each observation, and we always analyze these in tandem
with all LHB, MWH, and CXB component parameters tied together. Due to
systematic differences between the EMOS and EPN CCDs, particularly
with regard to details in particle background for extended sources, we
always fit these data separately and derive independent constraints on
sterile neutrino line emission from each set of detectors. Joint fits
would invite the introduction of a bias in the best fit model
parameters and could lead to the underestimation of errors by
artificially degrading the model goodness-of-fit (e.g., Baldi et
al. 2012).

Although the corresponding systematics are less severe we choose not
to coadd the datasets from the three separate observation intervals,
fitting them first separately and then simultaneously as we shortly
describe in detail. Our final results are taken from the latter. All
parameters are tested to see if they are well-determined, if fits are
sensitive to their values, and if limits on additional line emission
over a range of representative energies are significantly affected by
their variation. When thus justified, these are fixed as described
below. The following is based on extensive experimentation, from which
models with a minimum number of significant, variable parameters
emerge that lead to a robust limit on sterile neutrino line emission.

In the simultaneous fits, the Al $K\alpha$ and Si $K\alpha$ energies
and widths are fixed at the values determined from the individual
observation fits. The SWCX line energies are fixed at 0.56 and 0.65
keV, and their widths at 1 eV (negligible compared to the EPIC
resolution). The LHB temperature is fixed at 0.15 keV, the MWH
temperatures at 0.10 and 0.34 keV based on fits that included the RASS
spectrum, and the CXB slope at 1.46 \citep{kush02}. The EMOS1/EMOS2
offset is set to unity.

In what follows, we restrict discussion to the 100~pc radius circular
aperture spectral extraction region. The increase in counts provided
superior statistical accuracy to the 42.5 pc ($1.7\times$ the
half-light radius) region. Given the systematic mass uncertainties
(see below), we judge that this compensates for the formally smaller
uncertainty in the mass enclosed within the smaller aperture. While
the larger extraction aperture for the 150 pc region encompasses a
larger flux, the effect of vignetting increases the ratio of non-X-ray
to X-ray events and degrades the statistical accuracy.

We first consider the unbinned, unsubtracted\footnote{Here,
  ``background-subtracted'' and ``unsubtracted'' always refer to
  whether or not the QPB -- only -- is subtracted prior to fitting,
  and ``unbinned'' refers to the default energy intervals defined by
  the detector channel boundaries.}  spectra over the ``global''
  0.3-10 (0.4-4) keV energy range for the EMOS (EPN), fit by
  minimizing the modification of the C-statistic \citep{cas79}
  implemented in {\sc Xspec} as $cstat$ (we refer to this approach as
  ``$cstat$-$nbs$'', where nbs refers to ``no background
  subtraction''). Since the CXB is much smaller than the QPB, its
  normalization (relative to the all-sky average; see above) is also
  fixed at values determined from the separate observation fits. The
  SWCX norms are set to 0, and the relative observation-to-observation
  offsets are set to 1. Thus the free parameters are the separate SP
  broken power-law slopes for each detector and each observation, and
  the SP broken power-law energy break for each detector (linked at an
  identical value across observations). The free normalizations
  include those for the SP component and instrumental lines for each
  detector and each observation, and those for the LHB and MWH (tied
  for each set of EMOS or EPN detectors for each observation).

The following minor statistical issue arises with respect to the EPN
regarding out-of-time (OOT) events. The XMM-ESAS {\it pn-back} task
produces an OOT-subtracted spectrum that is not purely Poissonian, and
the $cstat$ statistic that we sometimes employ is strictly valid only
for data with Poisson errors. We address this by including an explicit
OOT background in these cases, scaling the OOT spectra created by the
XMM-ESAS {\it pn-spectra} task by 0.063 \citep{nml05}. Results using
the presubtracted spectra are consistent, as expected given the small
departure from Poissonian, with the approach we adopt -- although the
latter yields somewhat better fits.

Although these are not utilized to derive best-fits or sterile
neutrino line flux constraints, for purposes of illustrating the
relative strengths of the spectral model components and the overall
signal-to-noise of the data we show the coadded EMOS (EMOS1 and EMOS2,
all observations) and EPN (all observations) total spectra and mean
best-fitting model in Figure 1. 

\begin{figure}[ht]
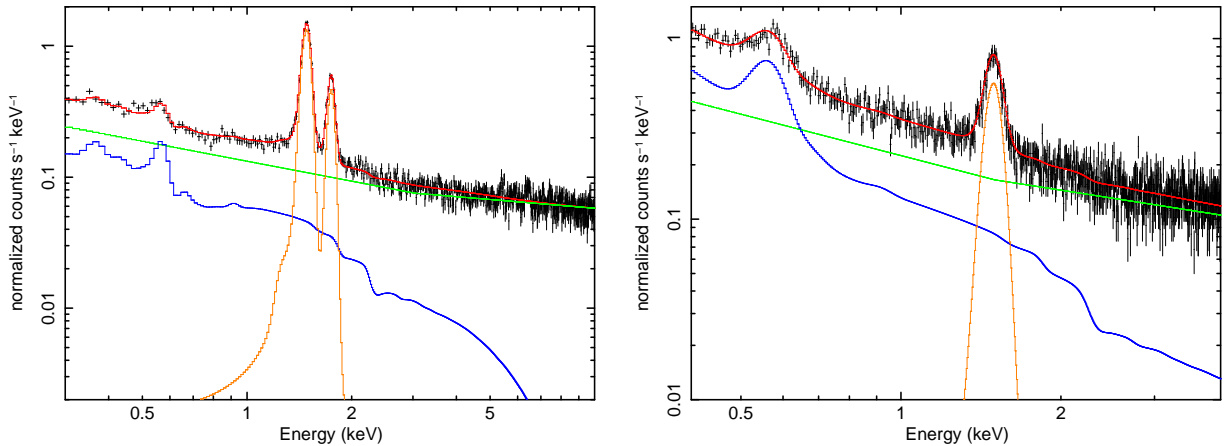

\includegraphics[scale=0.33,angle=-90]{mos_global_nbs_sscfit.ps}
\hfil
\includegraphics[scale=0.33,angle=-90]{pns_global_nbs_sscfit_unb.ps}
\caption{\footnotesize {The {\it left} panel (a) ({\it right} panel
    (b)) shows the total co-added EMOS (EPN) unsubtracted spectrum
    (black errorbars), and mean best-fitting model (red curve)
    composed of astrophysical (sum of LHB, MWH, and CXB; blue curve)
    background, particle background (SP; green broken power-law), and
    instrumental fluorescent lines (orange). This is meant to
    illustrate overall S/N for the unbinned spectra (note that the
    EMOS energy channels are wider than the EPN channels) and the
    spectral decomposition; co-added spectra are not used to derive
    spectral analysis results.}}
\end{figure}

Because of the prominence of the (unsubtracted) instrumental
fluorescent lines, deriving limits on additional line emission is
highly problematic in the 1.2-1.9 (1.3-1.65) keV energy range based on
EMOS (EPN) spectra. As a result it is sensible to split the spectrum
into low- and high-energy segments on either side, and treat these
separately. Moreover, since some of the model components are
negligible in one or the other of these segments one can apply simpler
models with parameters that are a subset of those in the global
fit. Ultimately, the constraints on sterile neutrino line emission are
derived from fits to these low- and high-energy sub-spectra, with the
role of global fits purely to provide a means of setting these up.

We proceed with the analysis described above in the low energy
spectral segments (0.3-1.2 keV for the EMOS, 0.4-1.3 keV for the EPN)
by freezing the instrumental line norms at the values derived from the
global fits, reactivating the SWCX lines (with normalizations tied
across observations), and with the SP component model converted to a
single-slope power-law allowed to vary from observation to
observation. We follow with fitting of the high energy (1.9-10 keV for
the EMOS, 1.65-4 keV for the EPN) spectral segments by freezing the
LHB, MWH, and SWCX normalizations and converting the SP component back
to a broken power law for the EMOS; parameters (slopes, break energies
if applicable) are tied across observations. Spectra and best-fit
models are shown in Figure 2.

\begin{figure}[ht]
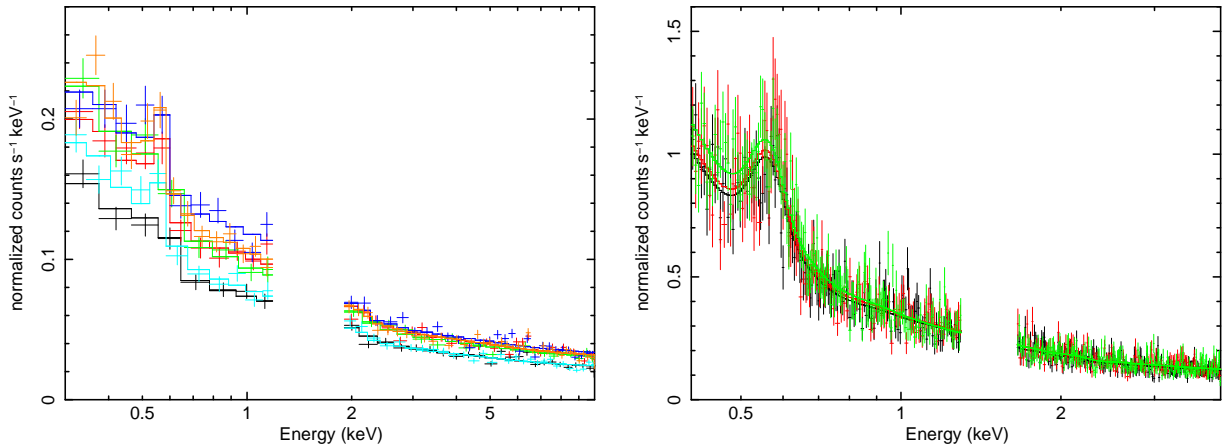

\includegraphics[scale=0.33,angle=-90]{mos12_hilo_nbs_sscfit_bin4_nolog.ps}
\hfil
\includegraphics[scale=0.33,angle=-90]{pn_hilo_nbs_sscfitC_bin2_nolog.ps}
\caption{\footnotesize{The {\it left} panel (a) ({\it right} panel
    (b)) shows the 6 (3) individual EMOS (EPN) spectra (with some
    binning) and best-fit models for the simultaneous $cstat$ fits to
    the unsubtracted spectra. Low- and high-energy segments, although
    shown together here, are separately fitted (see text for
    details).}}
\end{figure}

In our second approach we consider (QPB) background-subtracted, binned
(minimum 15 counts per bin) spectra fit by minimizing the $\chi^2$
statistic (we refer to this approach as ``$\chi^2$-$bs$''). The global
EPN bandpass may now be extended to 7 keV (residual instrumental
features at $\sim 4.5$ and $\sim 5.5$ keV that preventing this in the
unsubtracted spectrum are now subtracted out). With the increased
prominence of the CXB relative to the SP, we now fix the SP parameters
(though not their normalizations) to the best-fit values determined
above, and thaw the CXB norm (relative to its all-sky value) which is
tied across observations. For the low-energy sub-spectra, the
instrumental line normalizations are again fixed at the values derived
from the global fits and the SWCX lines are reactivated. In addition,
the CXB normalization is now fixed. In following this with fitting of
the high energy sub-spectra the LHB, MWH, and SWCX normalizations are
frozen, but the CXB normalization thawed and allowed to separately
vary for each observation. Again, for illustrative purposes, the
coadded subtracted spectra and corresponding mean best-fitting model
are shown in Figure 3. Spectra and best-fit models for the individual
spectra in each of the two energy intervals are shown in Figure 4.

\begin{figure}[ht]
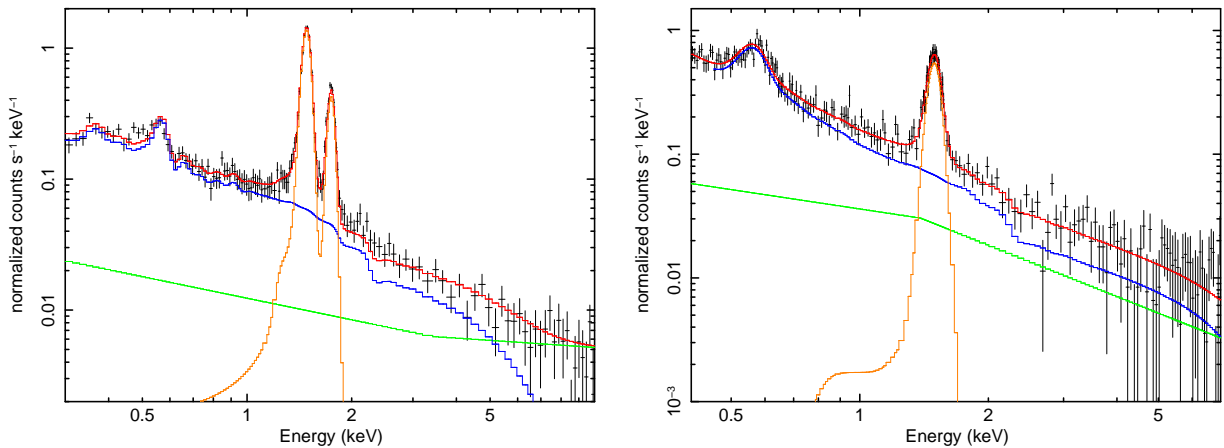

\includegraphics[scale=0.33,angle=-90]{mos_global_ssfit_bin2.ps}
\hfil
\includegraphics[scale=0.33,angle=-90]{pns_global_ssfit_bin2.ps}
\caption{\footnotesize{Same as Figure 1 for the subtracted spectra
  (with some binning).}}
\end{figure}

\begin{figure}[ht]
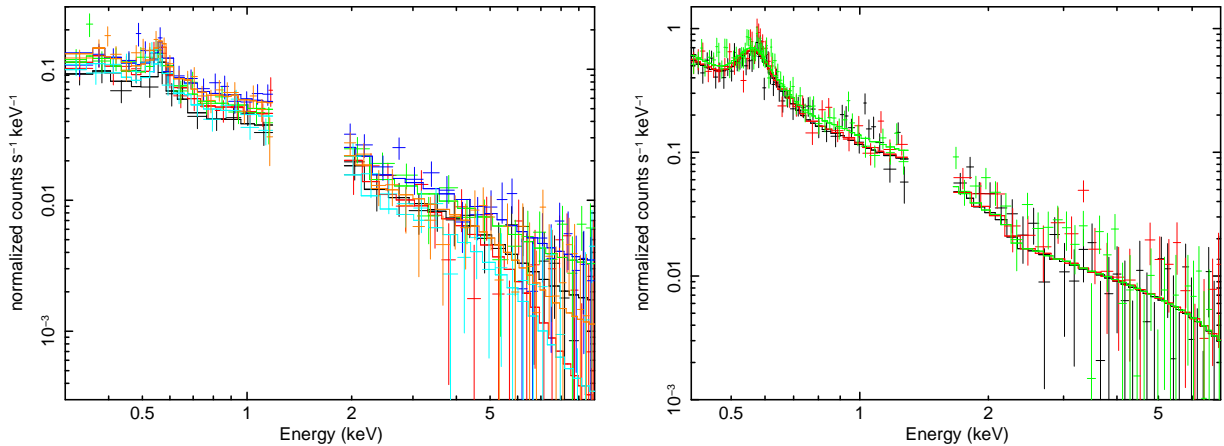

\includegraphics[scale=0.33,angle=-90]{mos12_hilo_ssfit_bin2.ps}
\hfil
\includegraphics[scale=0.33,angle=-90]{pn_hilo_ssfit_bin2.ps}
\caption{\footnotesize{Same as Figure 2 for the subtracted spectra
  (with some binning, and on a logarithmic scale).}}
\end{figure}

Finally, we also consider joint fitting of the total (unsubtracted)
and background (QPB) spectrum. Here the background spectrum, which is
not Poissonian, is binned to a minimum of 50 counts per bin and fitted
by application of $\chi^2$ statistics, while the total spectrum is
unbinned and fit by applying the $cstat$ statistic. Thus, the best-fit
model parameters are determined from minimization of a hybrid
statistic. The global fits proceed as in the first approach with SP
and instrumental parameters and normalizations tied in the models for
the total and QPB spectra, and the other (SWCX, LHB, MWH, and CXB)
component normalizations set to zero in the latter (thus no additional
parameters are introduced). For the low-energy sub-spectrum, the SP is
characterized by a broken power-law fixed across observations and
detectors. This is also the case for the high-energy EMOS spectrum,
while the SP is assumed to be a single-slope power-law for the
high-energy spectral EPN segment.

The fit statistics for all three approaches are displayed in Table 2,
along with the 90\% ($\Delta-statistic=2.71$, where ``statistic''
refers to the fit statistic -- either $\chi^2$, the modified
C-statistic, or the hybrid statistic described above) confidence upper
limits on the surface brightness of a narrow emission line with energy
fixed at 2.5 keV. Negative emission line fluxes are permitted when
deriving these limits.

\begin{deluxetable}{llllll}
\tabletypesize{\scriptsize}
\tablewidth{0pt}
\tablecaption{Fit Statistics -- 100~pc Extraction Region}
\tablehead{\colhead{method} & \colhead{bandpass} & \colhead{detector} & \colhead{stat} & \colhead{best-fit-stat} & $\Sigma_X(2.5)$}
\startdata 
$cstat$-$nbs$ & global & EMOS & $cstat$ & 3993/3841 & \nodata\\
      &    & EPN &     & 2108/2153 & \nodata\\
      & lo-en & EMOS &     & 340/337  & \nodata\\
      &    & EPN &     & 492/532  & \nodata\\
      & hi-en & EMOS &     & 3342/3225 & $-0.34_{-1.03}^{+0.99}$\\
      &    & EPN &     & 1416/1403 & $1.14_{-1.46}^{+1.51}$\\

$\chi^2$-$bs$ & global & EMOS & $\chi^2$ & 1545/2488 & \nodata\\
      &    & EPN &     & 2032/2025 & \nodata\\
      & lo-en & EMOS &     & 221/341  & \nodata\\
      &    & EPN &     & 499/500  & \nodata\\
      & hi-en & EMOS &     & 1021/1866 & $0.70_{-1.25}^{+1.26}$\\
      &    & EPN &     & 1354/1328 & $1.73_{-1.41}^{+1.48}$\\

hybrid   & global & EMOS & $hstat$ & 4359/4535 & \nodata\\
      &    & EPN &     & 2812/2505 & \nodata\\
      & lo-en & EMOS &     & 390/442  & \nodata\\
      &    & EPN &     & 654/681  & \nodata\\
      & hi-en & EMOS &     & 3630/3819 & $0.39_{-0.76}^{+0.80}$\\
      &    & EPN &     & 1732/1608 & $2.73_{-1.10}^{+1.46}$\\
\hline
\enddata 

\tablecomments{Shown are the fit statistics (per degree-of-freedom)
  for the best-fit model determined using the three analysis
  approaches -- unbinned, background-unsubtracted source spectral
  fitting with the (modified) C-statistic (``$cstat$-$nbs$''), binned,
  background-subtracted source spectral fitting with the $\chi^2$
  statistic (``$\chi^2$-$bs$''), joint unsubtracted-source/background
  spectral fitting using a combination of the $\chi^2$ and C-statistic
  (``hybrid'') -- see text for details. Also shown are the best-fit
  and 90\% confidence limits on the average surface brightness in
  units of $10^{-8}\ {\rm photons}\ {\rm cm}^{-2}\ {\rm s}^{-1} {\rm
  arcmin}^{-2}$, of a narrow emission line with energy fixed at 2.5
  keV.}
\end{deluxetable}

\subsection{Emission Line Flux Limits at 2.5 keV and Other Energies}

Having established the baseline models as described above, we derive
upper limits as in Loewenstein et al. (2009), and LK10. An unresolved
Gaussian component, stepped in 10 eV intervals over the relevant
bandpass, is added to the baseline model, and $\Delta-statistic=9.21$
(99\%) upper confidence levels on the line flux (that is permitted to
be negative) are computed. Limits for the low- and high-energy
segments are separately computed with the spectral model parameters
fixed or variable as described above -- except for the SWCX component
that we always fix at their best-fit values (with the -- necessary --
inclusion of this component in {\it XMM-Newton} spectra, limits
obtained in the $\sim$0.55-0.66 keV region using this method are
provisional). The limits, in 150 eV bins for the $\chi^2$-$bs$ and
$cstat$-$nbs$ approaches applied both to the EMOS and EPN detectors,
with zooms on the higher-energies for the latter, are shown in Figure
5. Given the similarity of the limits from the two approaches we
henceforth adopt the $\chi^2$-$bs$ limits. The $\chi^2$-$bs$
constraints are formally more restrictive for the EPN detector in the
4-7 keV bandpass. However as this is a result of an overall shift in
the allowed line flux and not of an improvement in accuracy and, given
the residual background artifacts in this energy range (\S 2.3.2), we
also restrict our subsequent discussion to constraints derived from
the EMOS detectors.

The expected line flux from sterile neutrino radiative decay in
Willman 1, produced by active-sterile neutrino transitions with
standard assumptions about the lepton asymmetry and thermal history of
the universe below $\sim$ 1 GeV (LK10 and references therein; also,
see below), is shown by the solid lines in Figures 5ab (the star in
(c) shows the prediction for $m_{\rm st}=2.5$~keV). The broken line
shows the allowed range taking the impact of hadronic uncertainties,
which affect the relation between the mass and the mixing angle, into
account \citep{als07}. An average total (sterile neutrino) dark matter
surface density of 200 $\rm M_{\odot}~{\rm pc}^{-2}$ is adopted (see
below). A sterile neutrino produced in this way in sufficient
abundance to compose all of the dark matter produces an emission line
that exceeds the observed limits if $m_{\rm st}> 5$~keV. That is, the
presence of a 2.5 keV line corresponding to the parameters of the
best-fit {\it Chandra} estimate is not confirmed. We further
illustrate this in Figure 6 that compares, for both approaches and for
the EPN and EMOS, measured spectra and best-fit models that include a
narrow 2.5 keV emission line with strength fixed at that predicted for
radiative decay of sterile neutrinos produced by non-resonant
oscillations. There is no evidence of such a feature in the {\it
  XMM-Newton} spectra.

\begin{figure}[ht]
\includegraphics[scale=0.25,angle=0]{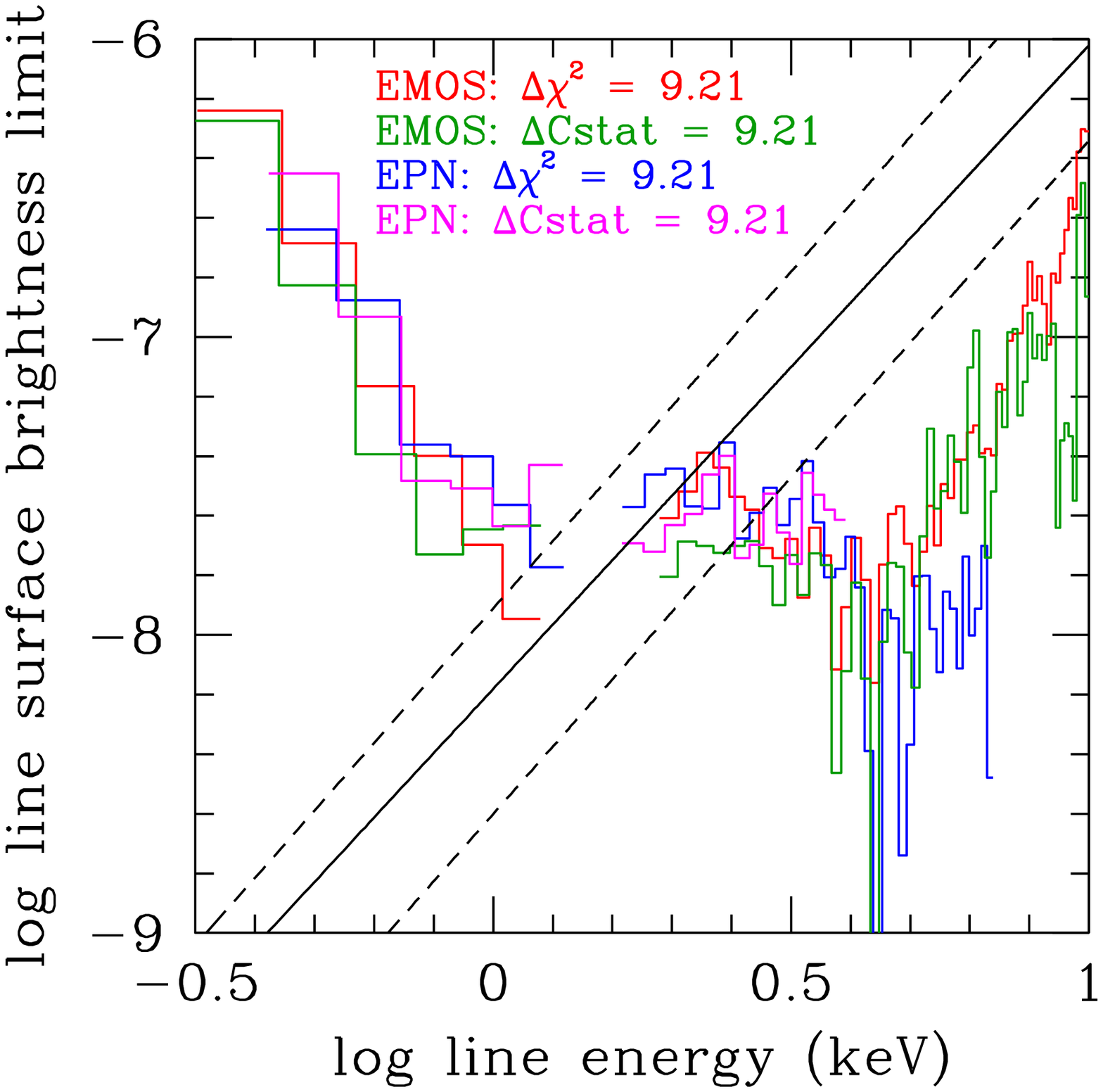}
\hfil
\includegraphics[scale=0.25,angle=0]{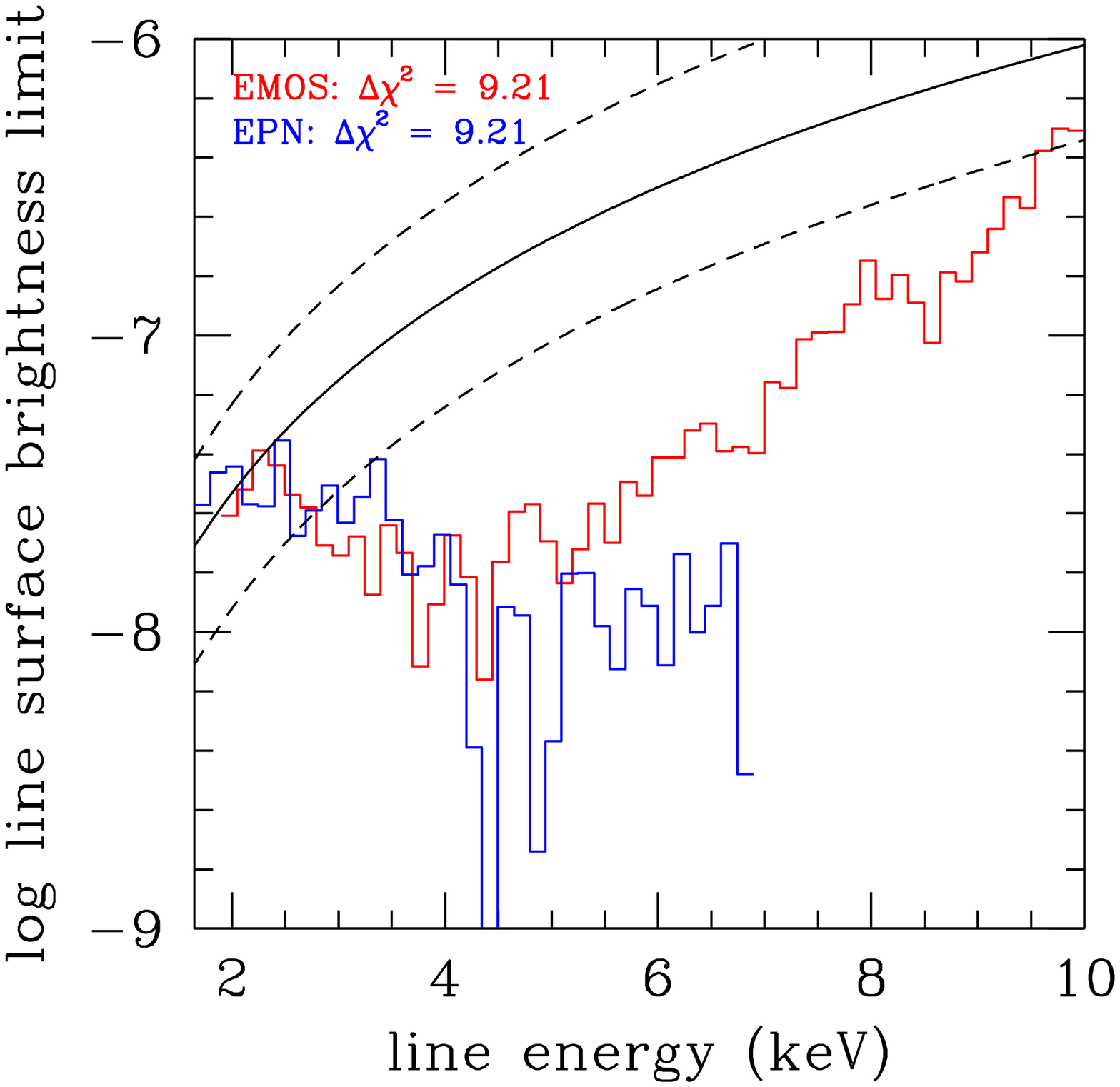}
\hfil
\includegraphics[scale=0.25,angle=0]{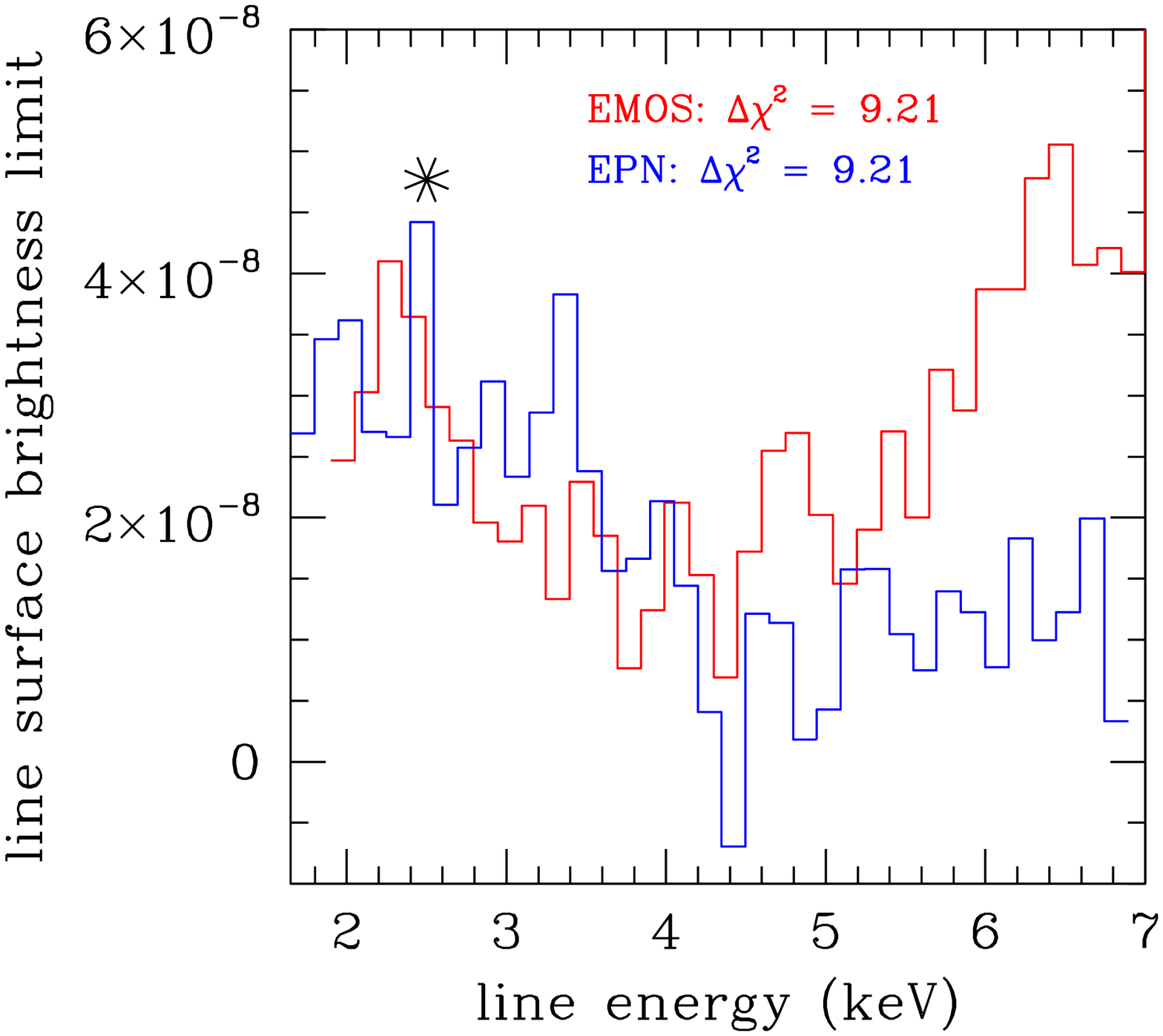}
\caption{\footnotesize{Upper limits on the emission line surface
    brightness, in units of ${\rm photons}\ {\rm cm}^{-2}\ {\rm
      s}^{-1} {\rm arcmin}^{-2}$ and averaged over the inner 100~pc of
    Willman 1, for both EMOS and EPN detectors and $\chi^2$-$bs$ and
    $cstat$-$nbs$ approaches as indicated in the embedded legend ({\it
      left} panel (a)), with focus on the high-energy spectra segment
    ($\chi^2$-$bs$ approach) in {\it middle} (b) and {\it right} (c)
    panels. The solid and broken lines show the expected line flux
    from sterile neutrino radiative decay in Willman 1, for an average
    total (sterile neutrino) dark matter surface density of 200 $\rm
    M_{\odot}~{\rm pc}^{-2}$ (see text), taking into account the
    hadronic uncertainties in production (Asaka et al. 2007). The star
    in (c) shows the mean prediction for $m_{\rm st}=2.5$~keV.}}
\end{figure}

\begin{figure}[ht]
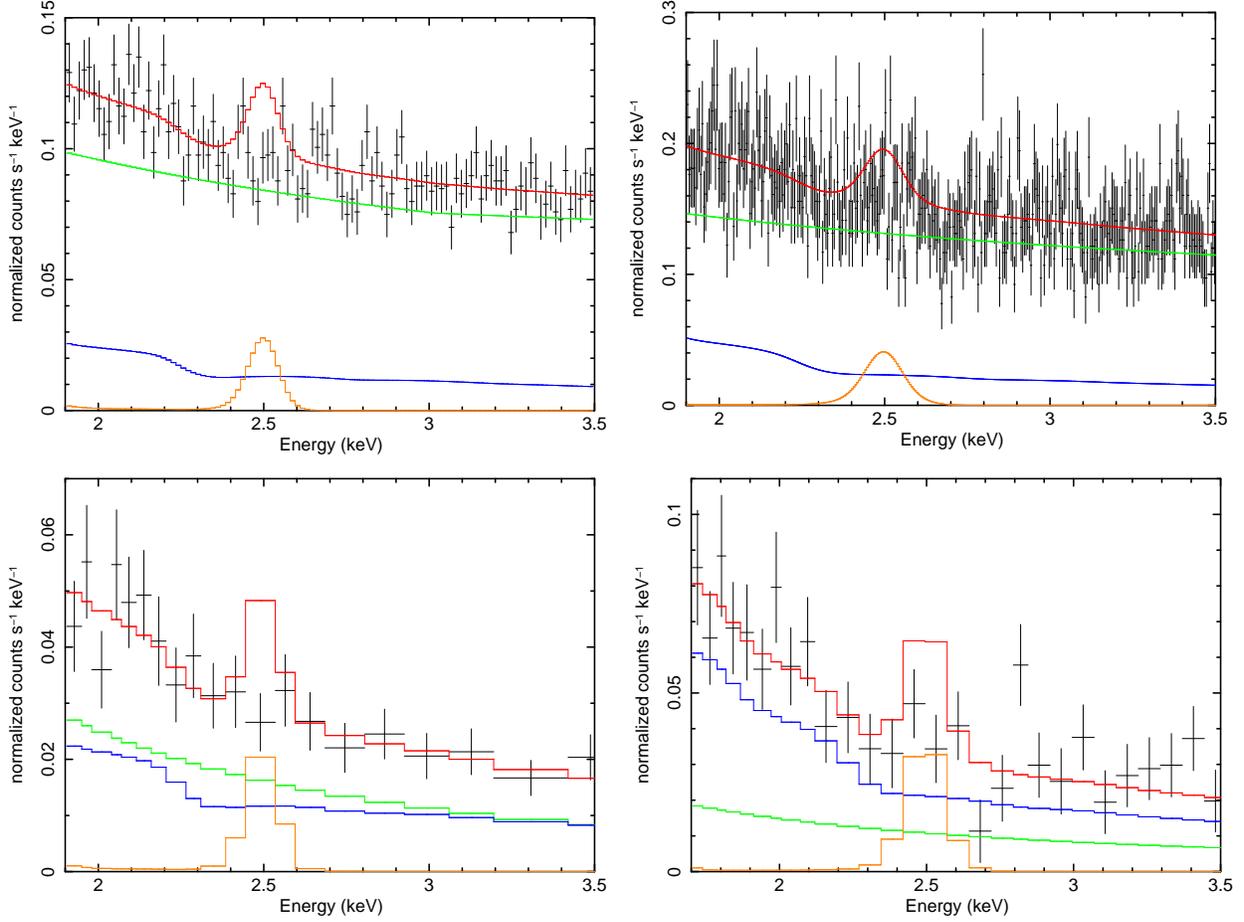

\includegraphics[scale=0.33,angle=-90]{mos_hi_nbs_sscfit_dwline_unb_nolog.ps}
\hfil
\includegraphics[scale=0.33,angle=-90]{pns_hi_nbs_sscfit_dwline_unb_nolog.ps}
\hfil
\vskip 0.1in
\includegraphics[scale=0.33,angle=-90]{mos_hi_ssfit_dwline_bin2_nolog.ps}
\hfil
\includegraphics[scale=0.33,angle=-90]{pns_hi_ssfit_dwline_bin2_nolog.ps}
\caption{\footnotesize{Unbinned EMOS ({\it upper left} panel (a)) and
    EPN ({\it upper right} panel (b)), and binned EMOS ({\it lower
      left} panel (c)) and EPN ({\it lower right} panel (d)) spectra
    (black errorbars) with best-fit models (and their decomposition)
    that include a narrow 2.5 keV emission line with strength fixed at
    that predicted for radiative decay of sterile neutrinos produced
    by non-resonant oscillations (see Figure 5). The upper and lower
    curves correspond to the $cstat$-$nbs$ and $\chi^2$-$bs$
    approaches, respectively. The color coding follows Figure 1, with
    the additional line component in orange.}}
\end{figure}

\section{Limits on Sterile Neutrino Parameters from the {\it XMM-Newton}
Spectrum of Willman 1}

The equations relating the dark matter projected surface mass density
$\Sigma_{\rm dm}$, X-ray observables (line energy $E_\gamma$, and line
surface brightness $\Sigma_{\rm line}$), and sterile neutrino
parameters (mass $m_{\rm st}$, mixing angle $\theta$, and fraction of
dark matter in sterile neutrinos $f_{\rm st}$), are the following
(see, e.g., LK10 and references therein):
\begin{equation}
m_{\rm st}=2E_\gamma,
\end{equation}
\begin{equation}
\Gamma_{\nu_s\rightarrow\gamma \nu_a}=5.52\times 10^{-32}\left(
\frac{\sin^2 \theta}{10^{-10}} \right) \ \left( \frac{m_{\rm st}}{\rm
  keV}\right)^5~{\rm s}^{-1},
\end{equation}
and
\begin{equation}
\Sigma_{\rm line}=3.95\times 10^{17}\Gamma_{\nu_s\rightarrow\gamma
  \nu_a}f_{\rm st} \left(\frac{\Sigma_{\rm dm}} {{\rm M}_{\odot}~{\rm
    pc}^{-2}}\right) \left(\frac{E_\gamma}{\rm keV}\right)^{-1} {\rm
  photons}\ {\rm cm}^{-2}\ {\rm s}^{-1} {\rm arcmin}^{-2},
\end{equation}
where the expression for the decay rate of relic keV sterile neutrinos
into an active neutrino and photon, $\Gamma_{\nu_s\rightarrow\gamma
 \nu_a}$, is given for Majorana sterile neutrinos.

As in LK10 we estimate the average dark matter surface mass density
within 100~pc based on projecting the best-fit NFW \citep{nfw97} mass
model in \cite{str08}; however, we now introduce a (tidal) truncation
radius of 930 pc \citep{s-c11} by adopting the ``$n=2$ BMO''
generalization of the NFW profile \citep{oh11}. This yields $135~{\rm
M}_{\odot}~{\rm pc}^{-2}$ from the dark matter in Willman 1, to which
we (conservatively) add $65~\rm M_{\odot}~{\rm pc}^{-2}$ (LK10)
associated with dark matter in the Milky Way halo to obtain a fiducial
total line-of-sight surface mass density $\Sigma_{\rm dm}(100~{\rm
pc})=200~\rm M_{\odot}~{\rm pc}^{-2}$. The resulting mass profile is
consistent with other estimates \citep{wol10,ae11}. However
$\Sigma_{\rm dm}(100~{\rm pc})$ is uncertain by a factor of 2 or more,
considering the errors associated with those in the velocity
dispersion profile, as well as the dynamical state of Willman 1 (see
discussion below). The form of the mass distribution is also uncertain
and, as would be expected for warm dark matter, there is evidence that
the dark matter density profile is flatter and smoother than might be
expected in a CDM universe on the relevant mass scale (Gilmore et
al. 2007; Cholis \& Salucci 2012 and references therein) and may even
be constant over 100 pc. The impact of these uncertainties on the {\it
average} density is smaller than that of these other sources (unlike
the case for dark matter annihilation calculations where the
emissivity depends on the square, rather than the first power, of the
density).

The upper limit on the dark matter radiative decay rate as a function
of energy from equation (3) with $f_{\rm st}=1$ is shown in Figure 7,
where we also show the limits for $\Sigma_{\rm dm}(100~{\rm
pc})=100~\rm M_{\odot}~{\rm pc}^{-2}$ (or, equivalently, $f_{\rm
st}=0.5$).  As in LK10 and Loewenstein et al. (2009) we map this into
excluded regions in the $m_{\rm st}$-$\theta$ sterile neutrino
parameter space under each of the following two separate assumptions:
(1) that all of the dark matter is composed of sterile neutrinos
produced by some unspecified mechanism, (2) that non-resonant
oscillations, as first suggested by \cite{dw94}, produced sterile
neutrinos at the abundance determined by $m_{\rm st}$ and $\theta$ as
calculated in Asaka et al. (2007) (Figure 8). The first region is
where $\Sigma_{\rm line}$, calculated from equations (1)-(3) assuming
$f_{\rm st}=1$, exceeds our inferred upper limits; the second where
the flux from sterile neutrinos produced at the minimal abundance from
oscillations exceeds these limits. The latter provides an absolute
constraint, assuming only a standard early thermal history of the
universe, since the oscillations cannot be turned off.

\begin{figure}[ht]
\includegraphics[scale=0.7,angle=0]{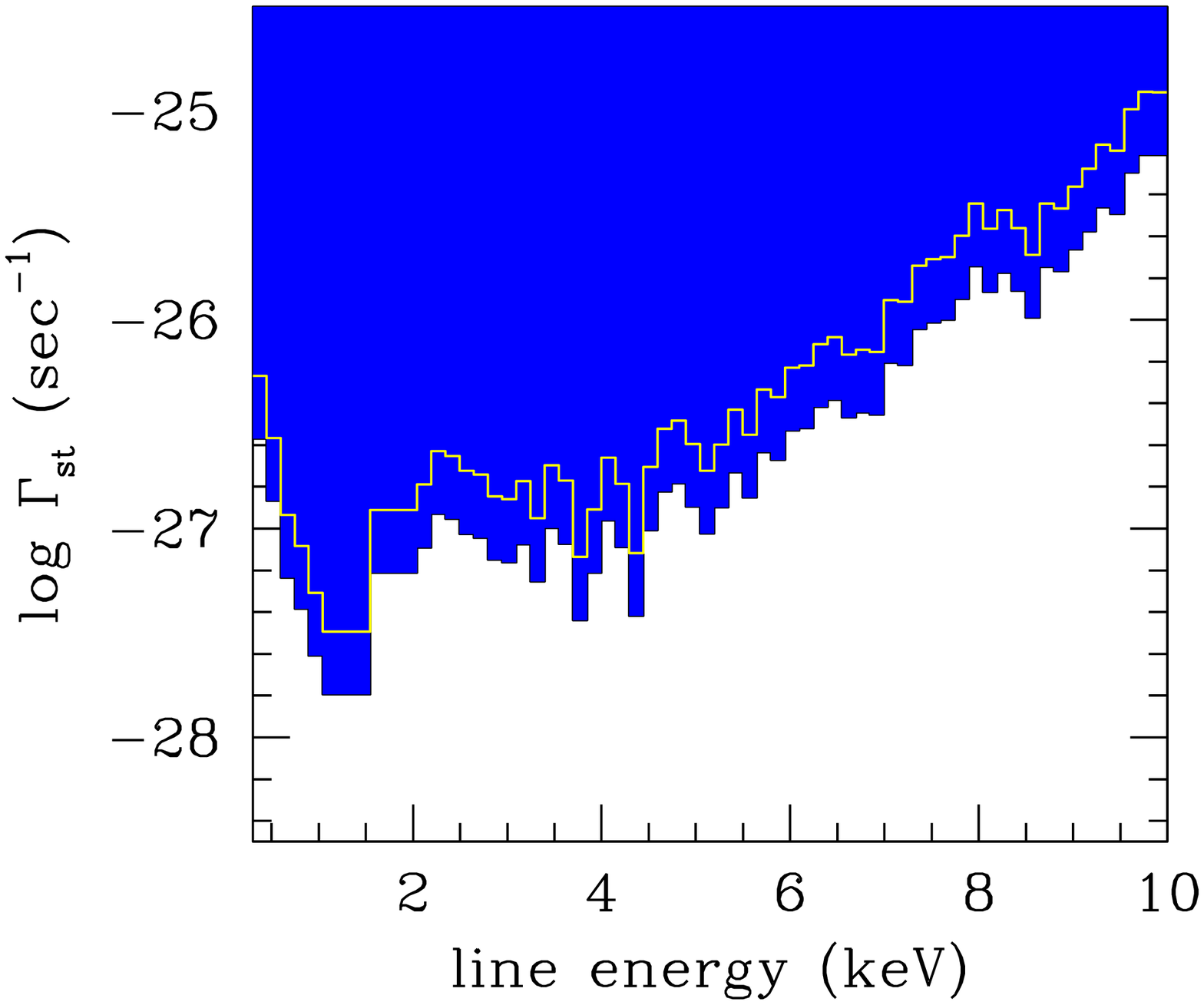}
\caption{\footnotesize{Excluded dark matter radiative decay rates
    (filled region) as a function of line energy derived from the
    upper limit on the average emission line surface brightness
    emerging from the inner 100~pc of Willman 1, assuming $\Sigma_{\rm
    dm}(100~{\rm pc})=200~\rm M_{\odot}~{\rm pc}^{-2}$ of decaying
    dark matter. The yellow histogram shows the corresponding limits
    for $\Sigma_{\rm dm}(100~{\rm pc})=100~\rm M_{\odot}~{\rm
    pc}^{-2}$ -- most of which would originate in the Milky Way
    halo.}}
\end{figure}

\begin{figure}[ht]
\includegraphics[scale=0.7,angle=0]{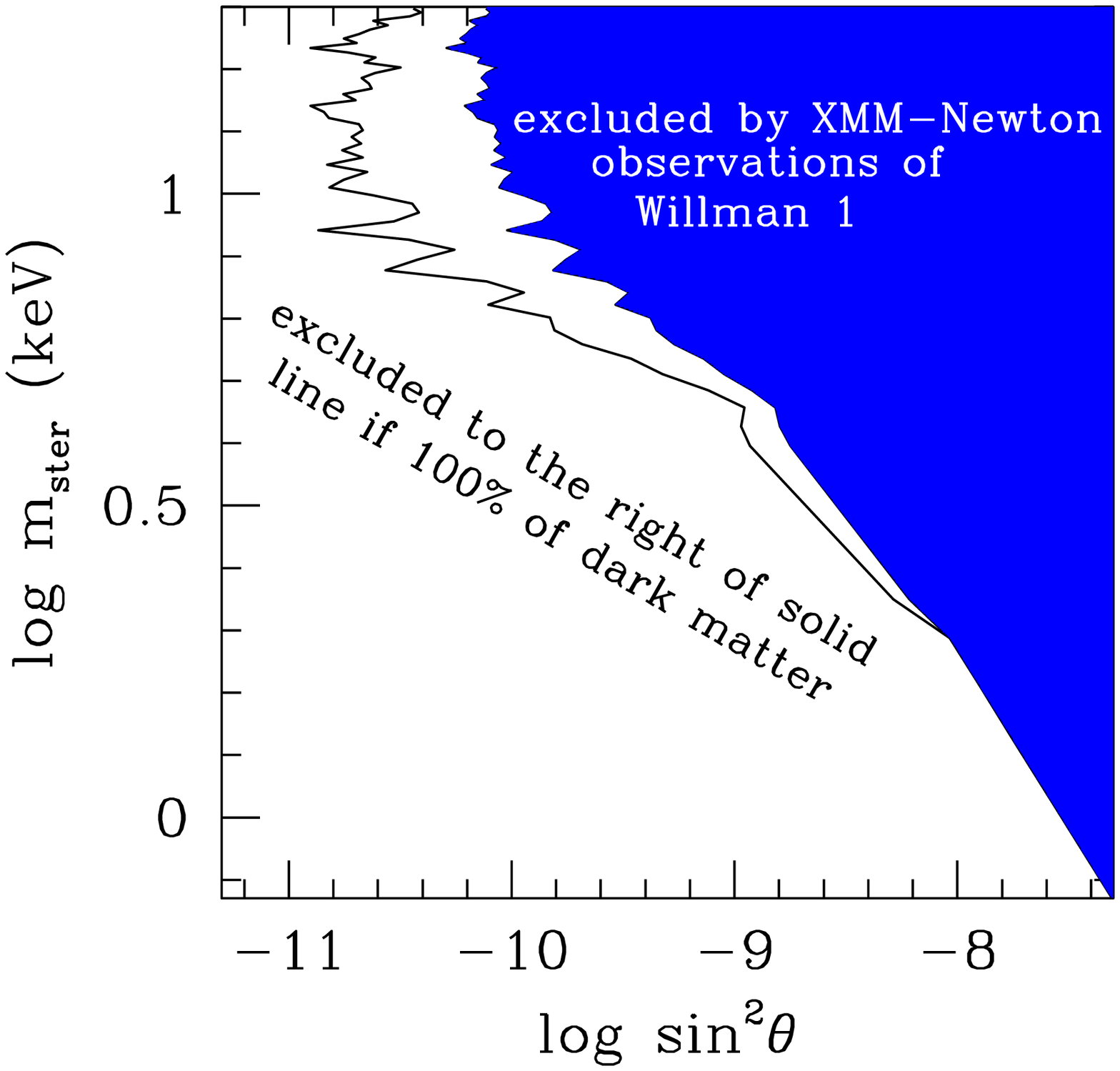}
\caption{\footnotesize{The shaded region in the $m_{\rm st}-\theta$
    sterile neutrino parameter plane is generally excluded assuming
    only the standard cosmological history below the temperature where
    production by neutrino oscillations occurs. The region to the
    right of the solid line is excluded if all of the dark matter is
    composed of sterile neutrinos produced by some (unspecified)
    mechanism.}}
\end{figure}

\subsection{Discussion and Conclusions}

We have derived upper limits on the radiative decay of dark matter
into keV photons in the ultra-faint dwarf spheroidal galaxy Willman 1,
and used these to place very general constraints on the mass and
mixing angle of any sterile neutrinos that contribute to the dark
matter. In doing so we adopted a total surface mass density (including
Milky Way dark matter in the line of sight) of $\Sigma_{\rm
  dm}(100~{\rm pc})=200~\rm M_{\odot}~{\rm pc}^{-2}$ in the solid
angle subtended by the {\it XMM-Newton} aperture corresponding to 100
pc at the Willman 1 distance of 38 kpc. While this estimate is based
on equilibrium dynamical models fit to the stellar velocity dispersion
profile, the dynamical status -- and even the very nature of Willman 1
-- were re-examined subsequent to our original X-ray observation of
Willman 1 with {\it Chandra}. Recent optical spectroscopy supports the
case that Willman 1 is (or was) a dwarf spheroidal and not a star
cluster; however -- in addition to the limited statistics of the
velocity dispersion profile -- doubts as to whether Willman 1 is in
dynamical equilibrium introduce a level of uncertainty in the Willman
1 dark matter mass that must now be estimated from non-equilibrium
dynamical modeling \citep{wil11}. Thus while our constraints are
formally comparable to previous limits based on dwarf spheroidal X-ray
spectroscopy (Loewenstein et al. 2009, Riemer-S\o rensen \& Hansen
2009, LK10, and references therein), we cannot claim to provide
additional new constraints on sterile neutrino parameters.

On the other hand we applied (for the first time) recently developed
extended-source analysis methods to demonstrate the robustness and
sensitivity of the {\it XMM-Newton} EMOS and EPN detectors to weak
line emission that may arise from dark matter radiative decay --
despite the fact that $\sim$half the data were discarded due to the
effect of flares. Clearly {\it XMM-Newton} has an important role to
play in such studies, with its larger effective area complementing the
lower and more stable background of {\it Suzaku} and the superior
spatial resolution (crucial for sources that, unlike dwarf
spheroidals, have prominent intrinsic discrete source emission) of
{\it Chandra}.

We find no confirmation of the {\it Chandra} evidence for an emission
line at 2.5 keV (Figures 5-6). Moreover, the abundance of $m_{\rm
st}=5$ keV sterile neutrinos produced by non-resonant oscillations has
recently been strongly proscribed based on {\it Chandra} imaging
spectroscopy of M31 \citep{wat11}, albeit in a region where baryonic
matter is still prominent. \cite{wat11} derive an upper limit $m_{\rm
st}<2.2$ keV for sterile neutrinos produced by this mechanism at the
abundance required to explain all of the dark matter ($f_{\rm
st}=1$). Combined with the lower bound from phase-space considerations
of $\sim 0.3$-0.4 keV \citep{tg79,ang10}, as well as those at higher
energy from the Milky Way and CXB \citep{ybw08}, this implies that
accounting for dark matter in this manner with sterile neutrinos may
remain viable only at masses where the decay line lies in a spectral
region dominated by blended emission features from the LHB and MWH --
and, thus, sensitivity to detection at X-ray CCD spectral resolution
is limited. As a result, the most significant expansion of the
searchable parameter-space boundary requires, above all, the sort of
leap in spectral resolution of diffuse sources possible with
microcalorimeter arrays such as the Soft X-ray Spectrometer that is
part of the {\it Astro-H}
observatory\footnote{http://heasarc.gsfc.nasa.gov/docs/astroh/}
scheduled for launch in 2013. As investigation of this lower $m_{\rm
st}$ regime proceeds, the region where off-resonant oscillations
produce pulsar kicks at the high end of the observed velocity
distribution \citep{fkmp03,kmm08} will be probed.

Alternatively, the sterile neutrino abundance may be subdominant
($f_{\rm st}<1$), or accounted for by means in addition to the DW
production mechanism. The production of sterile neutrinos via
oscillations may be resonantly enhanced in the presence of nonzero
lepton asymmetry \citep{sf99,kfs06,ls08}, thus opening up a larger
region in the $m_{\rm st}-\theta$ plane consistent with observed line
emission constraints and $f_{\rm st}=1$. Other suggested production
mechanisms, i.e. from inflaton or Higgs decay, do not involve
oscillations and are therefore independent of the mixing angle
\citep{kus06,st06,pk08}.  Furthermore, if sterile neutrinos are
produced above the electroweak scale \citep{kus06,pk08,kty10}, or if
some other sterile neutrinos decay and produce entropy
\citep{ask06,fkp09,fk11}, both the abundance and the clustering
properties of dark matter are affected.  However should the pulsar
kick and non-resonant oscillation domains be ruled out, the expected
detectability of sterile neutrino radiative decay lines becomes
essentially undetermined.

Despite ongoing dedicated production and direct detection experiments,
as well as intensive indirect searches, successful discovery of CDM in
the form of weakly interacting massive particles (WIMPs) remains
elusive. Given the prevalence of sterile neutrinos in extensions of
the standard model that explain the generation of neutrino masses, the
emergence of the keV scale as a natural one for some recently proposed
production scenarios \citep{kty10,mer12}, and indications that
cosmogonies dominated by WDM may explain discrepancies between CDM and
observations of small-scale structure and galaxy formation
\citep{lov11,mfl12,yc12}, strong physical motivation for 
their search persists. In addition, there are other decaying keV dark
matter candidates, one of which -- moduli dark matter -- we are
currently investigating \citep{lky12}.

\acknowledgments 

We thank Steve Snowden and Dave Davis for their advice on {\it
  XMM-Newton} data analysis issues, Beth Willman for comments on the
  draft manuscript and input on observing and funding proposals, and
  an anonymous referee for feedback. This work was supported by NASA
  ADAP Grant \#NNX11AD36G. AK acknowledges additional support from DOE
  Grant \#DE-FG03-91ER40662.

This work is based on observations obtained with {\it XMM-Newton}, an
ESA science mission with instruments and contributions directly funded
by ESA Member States and the USA (NASA), and utilized software
integrated and maintained at the {\it XMM-Newton} Science Operations
Center. Additionally, use was made of data and/or software provided by
the High Energy Astrophysics Science Archive Research Center
(HEASARC), which is a service of the Astrophysics Science Division at
NASA/GSFC and the High Energy Astrophysics Division of the Smithsonian
Astrophysical Observatory. Finally, use was made of NASA's
Astrophysics Data System, and the arXiv e-print service operated by
Cornell University.


{}

\clearpage


\begin{thebibliography}{}

\bibitem[Abazajian et al.(2001)]{afp01} 
Abazajian, K., Fuller, G.~M., \& Patel, M.\ 2001, \prd, 64, 023501

\bibitem[Abazajian et al.(2001)]{aft01} 
Abazajian, K., Fuller, G.~M., \& Tucker, W.~H.\ 2001, \apj, 562, 593

\bibitem[Amorisco \& Evans(2011)]{ae11}
Amorisco, N. C., \& Evans, N. W. 2011, \mnras, 411, 2118

\bibitem[Ando \& Kusenko(2010)]{ak10}
Ando, S., \& Kusenko, A.\ 2010, \prd, 81, 113006

\bibitem[Angus(2010)]{ang10}
Angus, G. 2010, \jcap, 3, 26

\bibitem[Asaka, Laine, \& Shaposhnikov(2007)]{als07} 
Asaka, T., Shaposhnikov, M., \& Laine, M. 2007, JHEP, 01, 091

\bibitem[Asaka \& Shaposhnikov(2005)]{as05}
Asaka, T., \& Shaposhnikov, M.\ 2005, Physics Letters B, 620, 17

\bibitem[Asaka et al.(2006)]{ask06} 
Asaka, T., Shaposhnikov, M., \& Kusenko, A.\ 2006, Physics Letters B,
638, 401

\bibitem[Asplund et al.(2009)]{agss09}
Asplund M., Grevesse N., Sauval A.J. \& Scott P. 2009, \araa, 47, 481

\bibitem[Baldi et al.(2012)]{bal12}
Baldi, A., Ettori, S., Molendi, S., Balestra, I., Gastaldello, F., \&
Tozzi, P. 2012, \aap, 537, 142

\bibitem[Bezrukov \& Shaposhnikov(2007)]{bs07}
Bezrukov, F., \& Shaposhnikov, M.\ 2007, \prd, 75, 053005

\bibitem[Boyanovsky(2008)]{boy08} 
Boyanovsky, D.\ 2008, \prd, 78, 103505

\bibitem[Boyanovsky \& Wu(2011)]{bw11}
Boyanovsky, D., \& Wu, J.\ 2011, \prd, 83, 043524 

\bibitem[Boyarsky, Ruchayskiy, \& Shaposhnikov(2009)]{brs09} 
Boyarsky, A., Ruchayskiy, O., \& Shaposhnikov, M.\ 2009, Annual Review
of Nuclear and Particle Science, 59, 191

\bibitem[Cash(1979)]{cas79}
Cash, W. 1979, \apj, 228, 939

\bibitem[Cholis \& Salucci(2012)]{cs12} 
Cholis, I. \& Salucci, P. 2012 (arXiv:1203.2954)

\bibitem[Dickey \& Lockman(1990)]{dl90}
Dickey, J., \& and Lockman, F. J. 1990, \araa, 28, 21

\bibitem[Dodelson \& Widrow(1994)]{dw94}
Dodelson, S., \& Widrow, L. M. 1994, \prl, 72, 17

\bibitem[Dunstan et al.(2011)]{dun11} 
Dunstan, R. M., Abazajian, K. N., Polisensky, E., \& Ricotti, M. 2011,
\prd, submitted (arXiv:1109.6291)

\bibitem[Ferrero et al.(2011)]{fer11} 
Ferrero, I., Abadi, M. G., Navarro, J. F., Sales, L. V., \& Gurovich,
S. 2011, \mnras, submitted (arXiv:1111.6609)

\bibitem[Fukugita \& Yanagida(1986)]{fy86}
Fukugita, M. \& Yanagida, T.~T., Phys.\ Lett.\ B, 174, 45 (1986).

\bibitem[Fuller et al.(2011)]{fk11} 
Fuller, G.~M., Kishimoto, C.~T., \& Kusenko, A.\ 2011, arXiv:1110.6479

\bibitem[Fuller et al.(2003)]{fkmp03} 
Fuller, G.~M., Kusenko, A., Mocioiu, I., \& Pascoli, S.\ 2003, \prd,
68, 103002

\bibitem[Fuller et al.(2009)]{fkp09} 
Fuller, G.~M., Kusenko, A., \& Petraki, K.\ 2009, Physics Letters B,
670, 281

\bibitem[Kishimoto, Fuller, \& Smith(2006)]{kfs06}
Kishimoto C. T., Fuller G. M., \& Smith C. J. 2006, \prl, 97, 141301

\bibitem[Kuntz \& Snowden(2008)]{ks08}
Kuntz, K. D., \& Snowden, S. L. 2008, \aap, 478, 575          

\bibitem[Kusenko(2006)]{kus06} 
Kusenko, A.\ 2006, Physical Review Letters, 97, 241301

\bibitem[Kusenko(2009)]{kus09}
Kusenko, A. 2009, Phys. Rept., 481, 1

\bibitem[Kusenko, Mandal, \& Mukherjee(2008)]{kmm08} 
Kusenko, A., Mandal, B. P., \& Mukherjee, A. 2008, \prd, 77, 123009

\bibitem[Kusenko \& Segr{\`e}(1997)]{ks97}
Kusenko, A., \& Segr{\`e}, G.\ 1997, Physics Letters B, 396, 197 

\bibitem[Kusenko, Takahashi, \& Yanagida(2010)]{kty10} 
Kusenko, A., Takahashi, F., \& Yanagida, T. T. 2010, Phys. Lett. B,
693, 144

\bibitem[Kushino et al.(2002)]{kush02}
Kushino, A., Ishisaki, Y., Morita, U., Yamasaki, N. Y., Ishida, M.,
Ohashi, T., \& Ueda, Y. 2002, \pasj, 54,327

\bibitem[Laine \& Shaposhnikov(2008)]{ls08}
Laine M., \& Shaposhnikov, M. 2008, \jcap, 6, 031

\bibitem[Loewenstein \& Kusenko(2010)]{lk10}
Loewenstein, M., \& Kusenko, A. 2010, \apj, 714, 652 (LK10)

\bibitem[Loewenstein, Kusenko, \& Biermann(2009)]{lkb09}
Loewenstein, M., Kusenko, A., \& Biermann, P. L. 2009, \apj, 700, 426

\bibitem[Loewenstein, Kusenko, \& Yanagida(2012)]{lky12} 
Loewenstein, M., Kusenko, A., \& Yanagida, T. T. 2012, in preparation

\bibitem[Lovell et al.(2011)]{lov11} 
Lovell, M., Eke, V., Frenk, C., Gao, L., Jenkins, A., Theuns, T.,
Wang, J., Boyarsky, A., \& Ruchayskiy, O. 2011, \mnras, in press
(arXiv:1104.2929)

\bibitem[Menci, Fiore, \& Lamastra(2012)]{mfl12}
Menci, N., Fiore, F., \& Lamastra, A. 2012, \mnras, in press
(arXiv:1201.1617)

\bibitem[Merle(2012)]{mer12} 
Merle, A., 2012, in Proceedings for the TAUP 2011 conference, Munich,
Germany, in press (arXiv:1201.0881)

\bibitem[Navarro, Frenk, \& White(1997)]{nfw97}
Navarro J. F., Frenk C. S., \& White S. D. M. 1997, \apj, 490, 493

\bibitem[Nevalainen, Markevitch, \& Lumb(2005)]{nml05}
Nevalainen, J., Markevitch, M., \& Lumb, D. 2005, \apj, 629, 172

\bibitem[Oguri \& Hamana(2011)]{oh11} 
Oguri, M., \& Hamana, T. 2011, \mnras, 414, 1851

\bibitem[Petraki(2008)]{pet08} 
Petraki, K.\ 2008, \prd, 77, 105004

\bibitem[Petraki \& Kusenko(2008)]{pk08} 
Petraki, K., \& Kusenko, A.\ 2008, \prd, 77, 065014 

\bibitem[Riemer-S\o rensen \& Hansen(2009)]{sh09}
Riemer-S\o rensen, S.; Hansen, S. H. 2009, \aap, 500, L37

\bibitem[S\'anchez-Conde et al.(2011)]{s-c11}
S\'anchez-Conde, M. A., Cannoni, M., Zandanel, F., G\'omez, M. E., \&
Prada, F. 2011, \jcap, 12, 011

\bibitem[Shaposhnikov \& Tkachev(2006)]{st06} 
Shaposhnikov, M., \& Tkachev, I.\ 2006, Physics Letters B, 639, 414

\bibitem[Shi \& Fuller(1999)]{sf99} 
Shi, X., \& Fuller, G.~M.\ 1999, Phys. Rev. Lett., 82, 2832

\bibitem[Snowden et al.(2008)]{sno08} 
Snowden, S. L., Mushotzky, R. F., Kuntz, K. D., \& Davis, D. S. 2008,
\aap, 478, 615

\bibitem[Strigari et al.(2008)]{str08} Strigari, L. E., Koushiappas,
  S. M., Bullock, J. S., Kaplinghat, M., Simon, J. D, Geha, M., \&
  Willman, B. 2008, \apj, 678, 614

\bibitem[Tremaine \& Gunn(1979)]{tg79} 
Tremaine, S., \& Gunn, J. E. 1979, \prl, 42, 407

\bibitem[Watson et al.(2011)]{wat11}
Watson, C. R.. Li, Z., \& Polley, N. K., \jcap, in press
(arXiv:1111.4217)

\bibitem[Willman et al.(2011)]{wil11}
Willman, B., Geha, M., Strader, J., Strigari, L. E., Simon, J. D.,
Kirby, E., Ho, N., Warres, A. 2011, \aj, 142, 128

\bibitem[Wilms, Allen, \& McCray(2000)]{wam00}
Wilms, J., Allen, A., \& McCray, R. 2000, \apj, 542, 941

\bibitem[Wolf et al.(2009)]{wol10} 
Wolf, J., Martinez, G. D., Bullock, J. S., Kaplinghat, M., Geha, M.,
Munoz, R. R., Simon, J. D., \& Avedo, F. F. 2009, \mnras, 406, 122

\bibitem[Yue \& Chen(2012)]{yc12} 
Yue, B., \& Chen, X. 2012, \apj, in press (arXiv:1201.3686)

\bibitem[Y\"uksel, Beacom, \& Watson(2008)]{ybw08} 
Y\"uksel, H., Beacom, J. F., \& Watson, C. R. 2008, \prl, 101, 121301

\end{thebibliography}
\end{document}